\documentclass[journal]{IEEEtran}
\usepackage{cite}

\usepackage{cite}
\usepackage{amsmath,amssymb,amsfonts}
\usepackage{algorithmic}
\usepackage{graphicx}
\usepackage{textcomp}
\usepackage{xcolor}
\usepackage{multirow}
\usepackage{empheq} % for making box
\newtheorem{theorem}{Theorem}

\def\BibTeX{{\rm B\kern-.05em{\sc i\kern-.025em b}\kern-.08em
    T\kern-.1667em\lower.7ex\hbox{E}\kern-.125emX}}

\newcommand{\vo}[1]{\boldsymbol{#1}}

\newcommand{\set}[1]{\mathcal{#1}}

\newcommand{\inner}[1]{\left\langle \vo{e}\phi_i\right\rangle}

\DeclareMathAlphabet{\mathbfsf}{\encodingdefault}{\sfdefault}{bx}{n}

\newcommand{\domain}[1]{\set{D}}

\begin{document}
\title{Modeling and Optimal Control of Hybrid UAVs with Wind Disturbance}

% \author{Michael~Shell,~\IEEEmembership{Member,~IEEE,}
%         John~Doe,~\IEEEmembership{Fellow,~OSA,}
%         and~Jane~Doe,~\IEEEmembership{Life~Fellow,~IEEE}% <-this % stops a space
% \thanks{M. Shell is with the Department
% of Electrical and Computer Engineering, Georgia Institute of Technology, Atlanta,
% GA, 30332 USA e-mail: (see http://www.michaelshell.org/contact.html).}% <-this % stops a space
% \thanks{J. Doe and J. Doe are with Anonymous University.}% <-this % stops a space
% \thanks{Manuscript received April 19, 2005; revised January 11, 2007.}}

\author{Sunsoo Kim$^{1}$, Niladri Das$^{2}$ and Raktim Bhattacharya$^{3}$
% \thanks{*This work was not supported by any organization}% <-this % stops a space
\thanks{$^{1}$Sunsoo Kim is a Ph.D student in the Department of Electrical and Computer Engineering,  Texas A\&M University, College Station, TX 77843, USA. Email: {\tt\small kimsunsoo@tamu.edu}}%
\thanks{$^{2}$Niladri Das is a Ph.D student in the Department of Aerospace Engineering, Texas A\&M University, College Station, TX 77843, USA. Email: {\tt\small niladridas@tamu.edu}}
\thanks{$^{3}$Raktim Bhattacharya is with the Faculty of Aerospace Engineering, Texas A\&M University, College Station, TX 77843, USA. Email:
        {\tt\small raktim@tamu.edu}}%
}

\markboth{Journal of \LaTeX\ Class Files,~Vol.~6, No.~1, January~2007}%
{Shell \MakeLowercase{\textit{et al.}}: Bare Demo of IEEEtran.cls for Journals}

\maketitle
\thispagestyle{empty}

\begin{abstract}
This paper addresses modeling and control of a six-degree-of-freedom unmanned aerial vehicle capable of vertical take-off and landing in the presence of wind disturbances. We design a hybrid vehicle that combines the benefits of both the fixed-wing and the rotary-wing UAVs. A non-linear model for the hybrid vehicle is rapidly built, combining rigid body dynamics, aerodynamics of wing, and dynamics of the motor and propeller. Further, we design an $\mathcal{H}_2$ optimal controller to make the UAV robust to wind disturbances. We compare its results against that of PID and LQR-based control. Our proposed controller results in better performance in terms of root mean squared errors and time responses during two scenarios: hover and level-flight.
\end{abstract}

% \fi
\IEEEpeerreviewmaketitle

\section{Introduction}
\IEEEPARstart{U}{nmanned} aerial vehicles (UAVs) have proved useful for both civil and military purposes \cite{singireddy2018technology}. Their popularity is increasing in applications such as surveillance, search and rescue operations, inspections, security, aerial photograph and video, mapping, and cargo system management \cite{army2010unmanned,canetta2017exploring,mazur2016pwc}.

Researchers and tech companies are developing different UASs to serve different purposes \cite{shakhatreh2019unmanned,liew2017recent}. We divide UAVs into two categories on the basis of their configurations: the rotary-wing UASs and the fixed-wing UASs. Rotary wing UAVs can take-off,land vertically, and hover at one position\cite{bolandi2013attitude}. While they need a small space for takeoff and landing, these UAVs can neither move fast nor fly long distances since they are not energy efficient. Compared to them, a fixed-wing of UAV is more power-efficient, hence it can fly for a longer duration of time and for further distance \cite{dorobantu2013system}. Despite these advantages, fixed-wing UAVs cannot take-off and land in small spaces because they need a runway to do so. Our proposed hybrid design aims to combine the advantages of the rotary-wing and the fixed-wing design.
%%%%%%%%%%%%%%%%%%%%%%%%%%%%%%%%%%%%%%%%%%%%%%%%%%%%%%%%%%

There are several hybrid UAV concepts \cite{saeed2018survey} such as a the dual system (combining fixed wing and rotary-wing), the tail-sitter, and the tilt-rotor. 
We classify these concepts according to their thrust direction. 
The simplest structure involves a dual system, which is a combination of two thrust directions: vertical and forward. 
In the tail sitter case, the heading of the vehicle is same as that of the thrust direction.
A tail sitter vehicle takes off vertically and then rotates pitch angle of body for the level flight. 
Unlike the tail-sitter type, in a tilt rotor/wing type of vehicle, it is the actuators that control the thrust direction. 
It takes-off, tilts the wing or rotor direction for level flight  \cite{yuksek2016transition,hanccer2010robust}, and lands vertically. 
For our research, we focus on the dual system type of UAV shown in Fig. \ref{Fig:configuration}. 
This is because the vehicle is mechanically simpler than the other hybrid UAVs and has the capability for VTOL and level flight. 
This UAV can take-off and land in smaller areas while having a large range of operation.

For the modeling of our hybrid UAV, we start with a conceptual design that satisfies our preliminary requirements.
First, we calculate the forces and moments coefficients on the wing using the vortex lattice method (VLM). 
After this aerodynamic analysis, we move on to the propulsion system. 
Here, we experimentally gather data on the thrust and torque from motor-propeller pair and generate a lookup table for our final model.
Next, we formulate the equations of motions based on rigid body dynamics. 
We use the detailed 3D model of our vehicle which includes properties like mass and inertia to complete our modeling. 
To perform simulations on this rigid body, we import the CAD (Computer Aided Design) model and lookup tables generated during propulsion analysis to SimScape \cite{Sims}. 
We exploit the built-in functionality of SimScape to import 3D design parameters and experimental data into the dynamic model of our UAV.

% Here, we can take advantage from Simscape software because any form of analysis like analytical math, 3D design parameters and experimental data, etc, can combined for one dynamics model of UAVs.

For UAV control, we mostly use the PID control method because of its ease of implementation \cite{li2012survey,salih2010modelling,xuan2019improved}. 
However, tuning PID gains to achieve the desired performance is a fairly challenging problem.
Experimental methods involving trial and error are used to tune these gains \cite{bo2016quadrotor, wang2016dynamics}. 
Thus, when UAVs encounter multiple uncertain stimuli such as wind gust, actuator noise, or just modeling errors, the controller may not work properly. 
Therefore, we need a more robust controller.
Researchers have developed adaptive control algorithms using model identification to handle uncertainties in the inertia and motor failure scenario \cite{schreier2012modeling,dydek2012adaptive}. 
They have also applied the robust control methods to handle the uncertainty in the system parameters like mass, inertia \cite{islam2014robust}, and actuator characteristics \cite{liu2015quaternion}.
However, there is little or no work on controller to reject wind disturbances with $\mathcal{H}_2$ control.
% and motor noise experienced during real-time flight, simultaneously.
Therefore, in this research, we focus on a robust optimal control of our hybrid UAV, which can reject wind disturbance.

The paper is organized as follows. In section \S \ref{Section:modeling}, we present modeling of our proposed hybrid UAV. Here, wing and thrust dynamics are presented in detail.  This is followed by the control algorithms, i.e. PID, Linear Quadratic Regulator (LQR), and $\mathcal{H}_2$ control in section \S \ref{section:Control}. In section \S \ref{section:result}, we introduce the simulation setup and show the results, followed by conclusions.
% The paper concludes with a few remarks and potential directions for further investigation.

%%%%%%%%
%% AIRCRAFT MODELNG
\section{Modeling of the Hybrid UAV} \label{Section:modeling}
% 1. Capability: speed range time weight 

% 2. Initial Configuration : depend on capability we can set design baseline
%     Wing area, ..
%     This parameters are updated with analysis of wing and thrust dynamics
    
% 3. Wing dynamics

% 4. Thrust dynamics

% 5. Non-linear model
     % Rigid body dynamics
     % wind disturbance

In this paper, we consider both fixed and rotary wing dynamics for our hybrid UAV.
We choose the flying wing shape, which does not have a tail wing as shown in Fig. \ref{Fig:configuration}. 
In this section, we are going to first discuss its design (its payload and flight characteristics), followed by its non-linear dynamics. 
A linearized dynamics model is also developed at the end of this section.
% To get optimal design, we essentially consider stability. actually, this aircraft does not have tail to get longitudinal stability, fly wing. Therefore, analysing stability in lateral and longitudinal direction is the first step depend on capability of aircraft. 
% In order to combine advantages and get rid of disadvantages of both vehicle configurations, a hybrid type of UAVs recently appeared. Those can be used in small areas and also in huge range thanks to combined advantages. This configuration can complete demanding applications today.
% There are many developed hybrid UAVs \cite{saeed2018survey} which are divided as the rotating arm, two-directional thrust, tail-sitter design. These concepts are classified according to their thrust and body direction. The first is the rotating arm type, that actuators control the thrust direction. It takes off and lands vertically and then tilts the wing or rotor direction to level flight \cite{yuksek2016transition}. The second is the two-directional thrust type. This takes off and lands with four propellers vertically and then shifting to level flight mode with one additional propeller. The third is the tail-sitter which has the same heading direction as the thrust direction. It takes off vertically and then rotates the whole body about the pitch direction for level flight \cite{verling2016full}. Further, we can see the flapping type, cyclo-copter type, and so on.
\subsection{Aircraft Design}
% \subsubsection{Capabilities}
Aircraft design is based on the desired capabilities we specify for our vehicle. 
Our aim is to develop a hybrid UAV which combines the advantages of both fixed wing and rotary wing type UAVs. 
The desired capabilities of the vehicle are set for a multi-functional application and are listed in Table \ref{table:capability}. 
They encompass that which is required broadly for applications such as drone deliveries, air surveillance and aerial photography, etc.
\begin{table}[h]
\caption{Vehicle desired capabilities} \label{table:capability}
\vspace{-0.3cm}
\centering
\renewcommand{\arraystretch}{1.5}
\begin{tabular}{|c|c||c|c|}
%\hline
%Capabilities & UAVs & Capabilities & UAVs & Capabilities & UAVs  \\
\hline
Type of operation & VTOL & Growth weight & 3.2 $kg$ \\
\hline
Flight time & 30 $min$ & Range & 3 $km$ \\
\hline
Level flight speed & 22 $m/s$  & Flight control & Auto Flight \\
% \hline
% Wing area & 16.76 & wing Span & 0.0750\\
\hline
\end{tabular}
\end{table}
% \subsubsection{Configuration} 
We start with an initial configuration. 
This configuration is able to sustain level flight, desired range, and satisfy payload characteristics.
% \comment{What do you mean by "set design baseline after considering ...the body structure and stability, etc."}
The final design of our UAV is selected after aerodynamic analysis of the initial configuration and through successive iterations of analysis.
% \comment{(iterations of what? a followed by b then a)}.

Aerodynamic stability analysis of the initial hybrid UAV configuration is an important step.
We used a numerical method called Vortex lattice method (VLM). 
This is a university-level technique used in computational fluid dynamics, which aids in the early stages of aircraft design. 
In this work, AVL (Athena Vortex Lattice) \cite{drela2004athena,melin2000vortex} and XFLR5 \cite{deperrois2009xflr5} softwares are used to implement VLM. 
This numerical method models a wing, the primary lifting surface, as an unbounded thin sheet of discrete vortices and calculates the induced drag and lift coefficients. 
It is also capable of calculating the air profile around an arbitrary wing with its rudimentary configuration alone.

For our UAV, we create batch codes and check the stability of our preliminary designs, followed by calculating forces and moments coefficients.
One can see in Fig. \ref{Fig:configuration} that our UAV does not have a tail wing, for the ease of manufacturing.
% \comment{you can talk about why you do not have a tail wing}
Hence, achieving longitudinal stability turns out to be the most challenging aspect of our design iterations. 
To address this problem, we select the re-flexed airfoil, Martin Hepperle (MH) 45 \cite{selig1996uiuc} and place the center of gravity (CG) in front of the neutral point (NP).
The optimal CG point is finally fixed. 
The corresponding level flight speed characteristics are shown in Table \ref{table:configuration}.
% Considering the stability of the aircraft, achieving longitudinal stability is the most challenging work because a flying wing does not have a tail wing.
% capabilities. we can set design baseline after considering the efficiency for level flight speed, flight range, payload, the body structure and stability, etc. And then final configuration of the vehicle is confirmed after analysis of dynamics on air vehicle. 
% % Here, we can set the values of 
% During the process for setting configuration, analysis of stability is the most important step. In this work, AVL (Athena Vortex Lattice) \cite{drela2004athena,melin2000vortex} and XFLR5 \cite{deperrois2009xflr5} software are used to implement the VLM method. Batch codes are created to check the stability, and calculate forces and moments coefficients. Considering the stability of the aircraft, achieving longitudinal stability is the most challenging work because a flying wing does not have a tail wing.

% Therefore, the re-flexed airfoil, Martin Hepperle (MH) 45 \cite{selig1996uiuc}, is selected and center of gravity (CG) is placed in front of the neutral point (NP). The optimal CG point and level flight speed is analyzed. In detailed wing configuration is shown in Table \ref{table:configuration}. 

For other payloads, we place the flight controller over the CG of the vehicle. 
The flight controller consists of an IMU (Inertial Measurement Unit) with integrated 3 axes accelerometer and gyroscope to measure accelerations and angular velocities.  
A telemetry radio for communication, RC receivers for manual controls, and a  6-cell LIPO battery for the power supply are placed in the vehicle. 
To ensure both hover flight and level flight, four propellers with a diameter of 9 inch and one propeller with a diameter 12 inch  are chosen, which are rotated by 1100 (kv) brush-less-electric motors. 
In the following subsection, we are going to first develop the rigid-body dynamics followed by modeling the wing dynamics and the thrust dynamics, which are then all combined to generate the full non-linear model for our proposed UAV. 

% \comment{Why two motors of different power rating? Where are they exactly placed. I see a forward flight propeller. What is the dimension of that propeller and what is the power rating of that motor ? What is MAC in the table. Can we mention the full-form of this ?}

% % and Fig. \ref{Fig:configuration}. 
% For the payload, the flight controller is placed at the CG (Center of Gravity) of the vehicle. This is composed of IMU (Inertial Measurement Unit) with integrated 3 axes accelerometer and gyroscope to measure accelerations and angular velocities. The telemetry radio for communication, RC receivers for manual controls and 6-cell LIPO battery for the power supply are placed  in the vehicles. To ensure both hover flight and level flight, four propellers with a diameter of 9 inch, 14 inch  are chosen, which are rotated by 1150 (W), 1350 (W) brushless-electric motors.
\begin{table}[h]
\caption{Wing configuration} \label{table:configuration}
\centering
\vspace{-0.3cm}
\renewcommand{\arraystretch}{1.5}
\begin{tabular}{|c|c||c|c|}
%\hline
%Configuration & UAVs & Configuration & UAVs & Configuration & UAVs\\
\hline
Wing span ($b$)  & 120 $cm$ & Wing area ($S$) & 3360 $cm^2$  \\
\hline
Root chord ($C_r$)  & 28 $cm$ & Mean Aerodynamics Chord  & 21.2 $cm$ \\
\hline
Tip chord ($C_t$) & 15 $cm$  &   $X_{CG}$  &  15 $cm$ \\%& Airfoil type  & MH 45 \\
\hline
Sweep angle & 25 $^\circ$  &  Height of winglet   & 15 $cm^2$\\
\hline
\end{tabular}
\end{table}

\subsection{Rigid body dynamics modeling} \label{sec:dyn_model}
\begin{figure}[ht]
\centering
\includegraphics[width=0.43\textwidth]{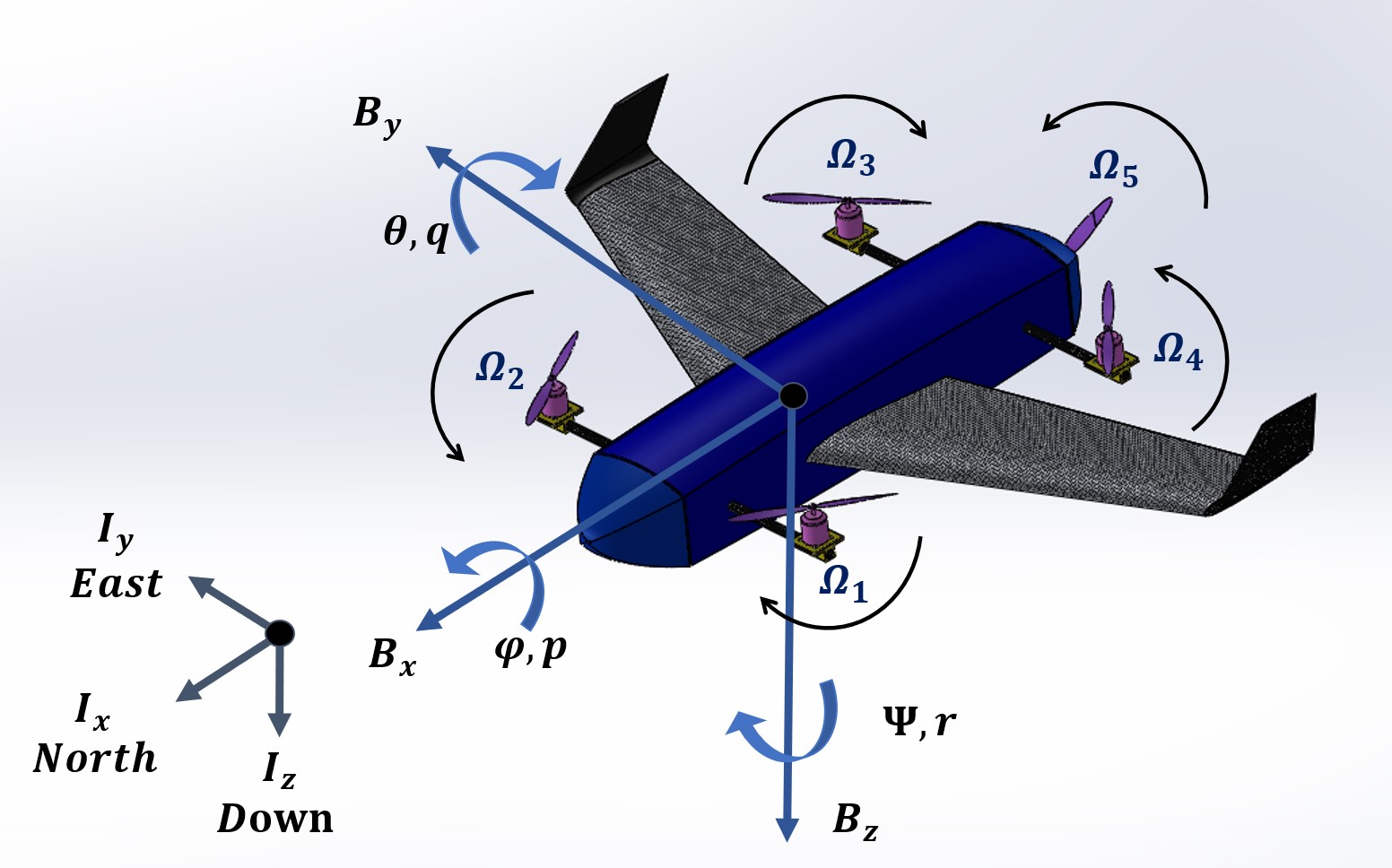}
\caption{Hybrid UAV configuration}
\label{Fig:configuration}
\end{figure}

We used Newton-Euler equations to develop the rigid body dynamics of the UAV. 
The 6-DoF dynamic model is shown in Fig. \ref{Fig:configuration} with the \textit{inertial frame} (I$_x$, I$_y$, I$_z$)  and \textit{body frame} (B$_x$, B$_y$, B$_z$) which follow the North-East-Down (NED) coordinate system. 
$\phi, \theta, \psi$ are the Euler angles in the inertial frame, and $p,q,r$ are angular velocities in the body frame about each axis. 
These 6 variables are the states for the \textit{rotational} motion of the UAV. 
Similarly, $x, y, z$ are the position in the inertial frame, and $u, v, w$ are velocities in the body frame about each axis. These 6 variables are states for \textit{translational} motion. Hence a total of 12 states of the vehicle dynamics are defined as
\begin{align*}
\setcounter{MaxMatrixCols}{20}
\vo{x}:= [ x \  y \  z \  u \  v \   w \  {\phi} \  {\theta} \  {\psi} \  p \  q \  r ]^T.
\end{align*}

\subsection{Wing dynamics modeling}
\begin{figure}[ht] 
\centering
\includegraphics[width=0.43\textwidth]{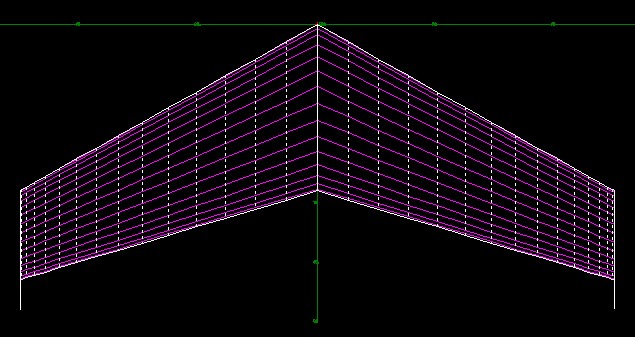}
\caption{The vortex lattice method panel}
\label{Fig:VLM}
\end{figure}

\begin{figure}[ht] 
\centering
% \includegraphics[width=0.32\textwidth]{lookuptableMx.png}
%\captionof{figure}{Some here}
\includegraphics[width=0.48\textwidth]{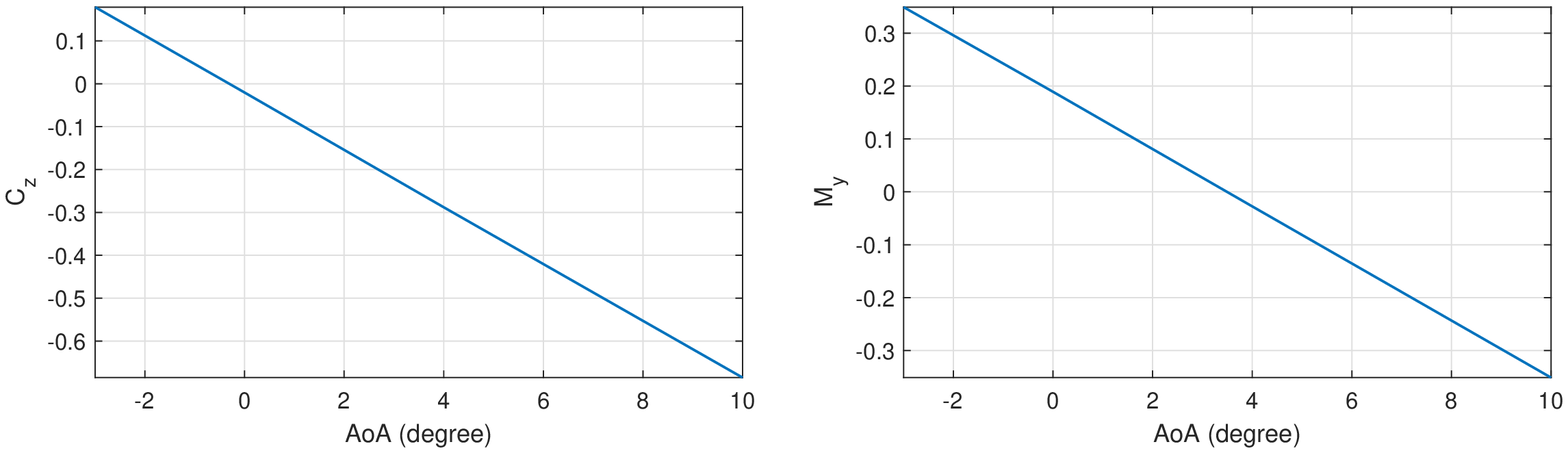}
\caption{Aerodynamic coefficient $C_Z$ and $M_Y$ : angle of attack varies from 0 to 10 $^\circ$. }
\label{Fig:wing_dynamics}
\end{figure}

The VLM is used to generate the aerodynamic coefficient of the wing body. 
The vortex lattice methods are based on solutions to Laplace’s Equation. 
Although VLM is a classical method in computational fluid dynamics, it can derive quite accurate results of aerodynamics for 3D Lifting surface, especially, in subsonic flow which we are concerning for modeling \cite{cummings2015applied}. 
The VLM calculations are mainly processed with the boundary condition and Kutta-Joukowski theorem \cite{anderson2015aerodynamics}. 
The wing is discretized to small panels as Fig. \ref{Fig:VLM}. 
Vortices are placed on each panel and the corresponding strength $\Gamma_i$ is obtained to satisfy the boundary condition theorem. 
Finally, forces and moments are computed by the Kutta-Joukowski theorem, which are presented as
\begin{subequations}
\begin{align}
\centering
L_i &= \rho V_\infty \times \Gamma_i \Delta b_i  \quad \text{(Lift of the panel i)},\\ \quad L &= \sum_{i=1}^{N} L_{i} \quad \text{(Lift of the Wing)} \label{eqn:Lift} \\
D_i &= \rho V_\infty  \times  \Gamma_i \Delta b_i \quad\text{(Drag of the panel i),}  
\quad\\  D &= \sum_{i=1}^{N} D_{i} \quad \text{(Drag of the Wing)} \label{eqn:Drag}
\end{align}
\end{subequations}
% \begin{figure}
% \centering
% \includegraphics[width=.9\textwidth]{AVL_vortex.png}
% \caption{Discretized panel(Left) and horse shoe vortex on each panel(Right)}
% \label{AVL}
% \end{figure}
% \begin{align}
% \centering
% \label{L}
% &&L_i &= \rho V_\infty \times \Gamma_i \Delta b_i&&\text{Lift of panel i}, \hspace{.5cm}
% &&L = \sum_{i=1}^{N} L_{i}&&\text{Lift of Wing}\\
% &&D_i &= \rho V_\infty  \times  \Gamma_i \Delta b_i&&\text{Drag of panel i} \hspace{.5cm}
% \label{D}
% &&D = \sum_{i=1}^{N} D_{i}&&\text{Drag of Wing}
% \end{align}
where, $\rho$ is the air density, $V_{\infty}$ is the free stream velocity, $\Gamma_i$ is the vortex strength in panel $i$, and $b$ is the length of the vortex segment along the quarter-chord line.
% \comment{A picture of the vortex panel will be of great help to the readers.}
The AVL software is used to obtain the aerodynamic variables of the wing. 
The result sets, which depend on seven input variables, are made up of a look-up table. 
The seven input factors are as follows: angle of attack, side slip angle, roll/pitch/yaw rate, elevator, and aileron deflection angle. 
One of the aerodynamic results from AVL is shown in Fig. \ref{Fig:wing_dynamics}. 
The resulting coefficients are then used to calculate the forces and moments for each body axis using 
\begin{subequations}
\begin{align}
    F_x = q_{\infty} S C_{F_y}, \  F_y = q_{\infty} S C_{F_y}, \  F_z = q_{\infty} S C_{F_z},  \\
    M_x = q_{\infty} S C_{M_x}, \  M_y = q_{\infty} S C_{M_y}, \ M_z = q_{\infty} S C_{M_z}, 
\end{align}
\end{subequations}
where $q_{\infty}$the dynamic pressure is  $q_{\infty} = \frac{1}{2} \rho V_{\infty}^2$.

% These aerodynamic forces and moments act on the CG of the vehicle.
% where $S$ is the reference wing area, $b$ is the reference wingspan, $c$ is the reference chord length, and $q_{\infty}$the dynamic pressure which is expressed $q_{\infty} = \frac{1}{2} \rho V_{\infty}^2$. These aerodynamic forces and moments are then act on the CG of the vehicle.

\subsection{Thrust dynamic modeling}
\begin{figure}[ht] 
\centering
\includegraphics[trim=0cm 0.8cm 0cm 0cm,width=0.48\textwidth,height=5cm]{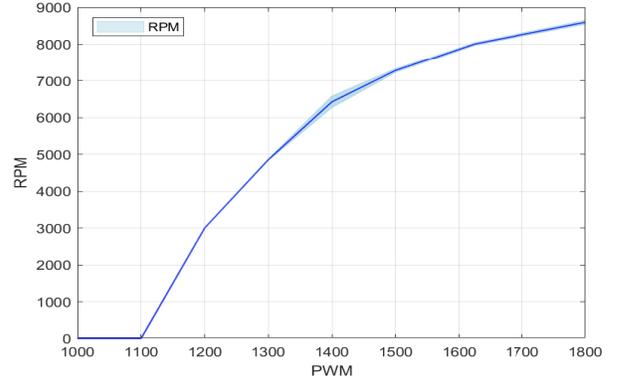}
\caption{The motor RPM result from the experiment with motor} %(e-flite power 10)} 
\label{Fig:Motor}
\end{figure}

\begin{figure}[ht] 
\centering
\includegraphics[trim=0cm 0.8cm 0cm 0cm,width=0.48\textwidth,height=5cm]{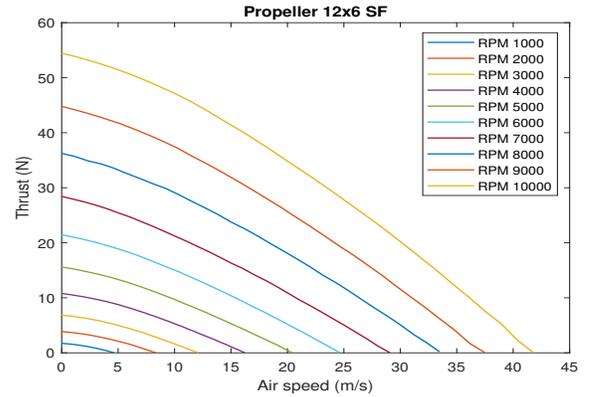}
\caption{The propeller thrust from the propeller performance data}
\label{Fig:Prop}
\end{figure}

Since the hybrid UAV is intended to perform level flights, free stream velocity should be considered when the thrust and torque of propellers are derived. 
Conventionally, DC motor parameter identification and blade element theory \cite{seddon2011basic} are  applied to get dynamic model. 
However, for more accurate modeling, we use the experimental method to derive brushless DC motor and propellers performance data, wind tunel test data \cite{propdata}, and generate lookup tables. 
The result of experiment on brushless DC motor with varying pulse width modulation (PWM) signal input is shown in Fig. \ref{Fig:Motor}. 
The results of thrust and torque from the propeller 12 $\times$ 6 SF (Slow Flight) that depend on wind velocity acting on the wing (free stream velocity) and RPM of motor are shown in Fig \ref{Fig:Prop}. 

% For the brevity of the paper, other results (plots)  of thrust and torque of propeller 10x5 is disregarded.

\subsection{Final non-linear model}

\begin{figure*}[t] 
\centering
\includegraphics[width=\textwidth]{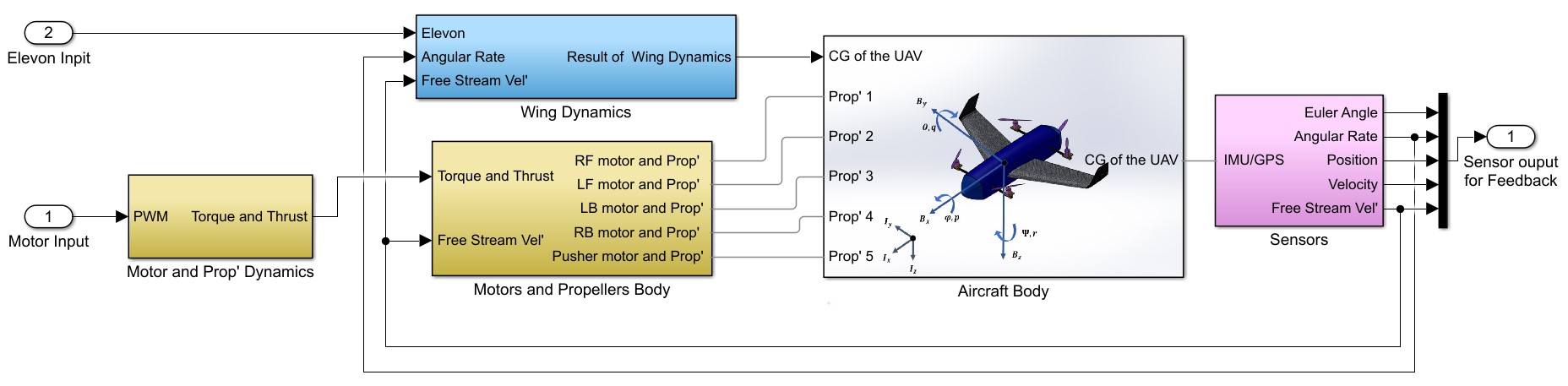}
\caption{6-DOF non-linear simulation of the hybrid UAV}
\label{Fig:Non_model}
\end{figure*}  
Our hybrid vehicle is developed as a 3D model using CAD.
% 3D design of the vehicle is developed computer-aided design (CAD) with the final configuration. 
This 3D model which include mass, inertia, and coordinate information is imported to Simscape software in Simulink \cite{Sims}.
% and build the rigid body of the hybrid UAV. 
The final non-linear 3D model is constructed by combining wing, motor, and propeller dynamics which are previously discussed as shown in Fig. \ref{Fig:Non_model}. This is the rapid modeling representing the equations of motion of UAVs.
 
% \begin{figure*}[t] 
% \centering
% \includegraphics[width=\textwidth]{simulink_image.jpg}
% \caption{6-DOF non-linear simulation of the hybrid UAV}
% \label{Fig:Non_model}
% \end{figure*}  

\subsection{Linearized model} \label{sec:linearized_sys}
We linearize the non-linear model of our hybrid UAV. 
Our aim is to design the controller for the \textit{attitude control} during level flight and during hovering. 
The following linear model is used in designing a $\mathcal{H}_2$ optimal control to make the system robust to wind gusts.
% and motor noises.
Since we are only interested in attitude control, we consider the corresponding state space $ [\theta \ u \  w \  q ]^T$ for level flight (Longitudinal motion) and  $  [\phi \ \theta \ \psi \ p \ q \ r ] ^T $ for hover flight. 
We calculate the linearized dynamics separately for level flight and hovering. 
For level flight, trim states are: 
\begin{align}
    p = q = r = 0, \quad V_{\infty} = \text{22.49} m/s,\label{eqn:trim_level}
\end{align} 
and for hovering, trim states are
\begin{align}
    p = q = r = 0, \quad u=v=w=0.\label{eqn:trim_hover}
\end{align}
% \comment{Can we write the values of the trim states?}.
The linearized error dynamics about the trim points are modeled as
% \vspace{-0.4cm}
\begin{subequations} \label{eqn}
\begin{align}
     \Dot{\vo{x}}(t) &= \vo{A} \vo{x}(t) + \vo{B}_{u} \vo{u}(t) + \vo{B}_w \vo{w}(t),\\
     \vo{y}(t) &= \vo{C} \vo{x}(t), 
 \end{align}
\end{subequations}
with states
$ \vo{x}:=  [\delta\theta \ \delta u \ \delta w \ \delta q ]^T$ for level flight and $\vo{x}:=  [\delta\phi \ \delta\theta \ \delta\psi \ \delta p \ \delta q \ \delta r ] ^T $ for hovering, which are perturbations on states about trim point. The system matrices for level flight are 
% for level flight trim state as
\begin{align} \label{eqn:sys_level}
&\vo{A}
    =\left[\begin{smallmatrix}
0 & 0 & 0 & 1 \\
0 & 0.0002 & -0.0235  & -0.1360\\
0 & 0.0011 &  -0.1793 &  20.4845\\
0 & 0.0135 &  -2.1745 &  -3.2657
\end{smallmatrix} \right], \ \vo{B}_u =\left[\begin{smallmatrix}0 \\ 0.0009 \\ -0.0407 \\ -0.6544 \end{smallmatrix}\right], \nonumber \\
&\vo{B}_w =\left[\begin{smallmatrix} 0 \ 0 \ 0 \ 1 \end{smallmatrix}\right]^T, \ \vo{C} = I_{4 \times 4},\end{align}where $u$ is elevon deflection ($\delta_e$). And for hover flight 
\begin{align}\label{eqn:sys_hover}
&\vo{A}
    =\left[\begin{smallmatrix}
        \vo{0}_{3 \times 3 } & \vo{I}_{3 \times 3 }\\
        \vo{0}_{3 \times 3 } & \vo{0}_{3 \times 3 }
\end{smallmatrix} \right], \ \vo{B}_u =\left[\begin{smallmatrix} \vo{0}_{3 \times 4 }\\ -153.5	\ 153.5\	153.5\	-153.5\\
36.9\	-37.1\quad	36.9\	-37.1\\
-1.8\quad	-1.8\quad	1.8\quad	1.8 \end{smallmatrix}\right], \nonumber
\\
&\vo{B}_w 
=\left[\begin{smallmatrix} 0 \ 0 \ 0 \ 1 \ 1 \ 1 \end{smallmatrix}\right]^{T} , \ \vo{C} = I_{6 \times 6},\end{align}
where $u$ is four motor input ($PWM_i,i=1 \sim 4$).

\section{Control}\label{section:Control}
% \indent Regarding about the control for the hybrid UAVs, the PID control method mostly used in the market because it is more simple than other controllers to start. But, the designed PID gains may not be worked in the uncertainties and disturbances. Also, the gain tuning is not easy due to the structure becoming more complicated. Thus, when UAVs encounter multiple uncertain stimuli such as modeling parameters, wind gust, noise on the actuator and so on, we require more robustness. As such, we demand a more optimal and robustness controller.
In this paper, we present a $\mathcal{H}_2$ optimal controller for our proposed novel hybrid UAV. 
Our hybrid vehicle harnesses the advantages of both fixed wing and rotor wing UAVs. 
% \subsection{Linear dynamic system}
We consider the following linear system which models the error dynamics about the trim points
\begin{subequations}\label{eqn:lin_system}
\begin{align}
\Dot{\vo{x}}(t) &=  \vo{A} \vo{x}(t) + \vo{B}_w \vo{w}(t) + \vo{B}_u \vo{u}(t),\\
\vo{z}(t) &= \vo{C}_z \vo{x}(t) + \vo{D}_u \vo{u}(t),\\
\vo{y}(t) &= \vo{C}_y\vo{x}(t),
\end{align}
\end{subequations}
where $\vo{x} \in \mathbb{R}^{n_x}$, $\vo{y} \in \mathbb{R}^{n_y}$, $\vo{z} \in \mathbb{R}^{n_z}$ are the state vector, the measured output vector, and the output vector of interest, respectively. Variables $\vo{w} \in \mathbb{R}^{n_w}$ and $\vo{u} \in \mathbb{R}^{n_u}$ are the disturbances and the control vectors, respectively. 

We are interested in designing a full state feedback $\mathcal{H}_2$ optimal controller for the system in Eq. \ref{eqn:lin_system}, i.e.,
\begin{align} \label{eqn:control_law}
\vo{u}(t) = \vo{K} \vo{x}(t),
\end{align}
such that the closed loop system is stable and the effect of the disturbance is attenuated to a desired level.
We perform a comparative study of the performance of the $\mathcal{H}_{2}$ optimal control with that of the conventional PID control and the LQR, when applied to our system. $\mathcal{H}_{2}$ control is expected to achieve better control performance in presence of disturbances since it incorporates the disturbance term $\vo{B}_{w}$ inside the  optimization process. 
Now, we briefly discuss the three controllers.

\subsection{PID controller}
\begin{figure}[ht] 
\centering
\includegraphics[width=0.48\textwidth]{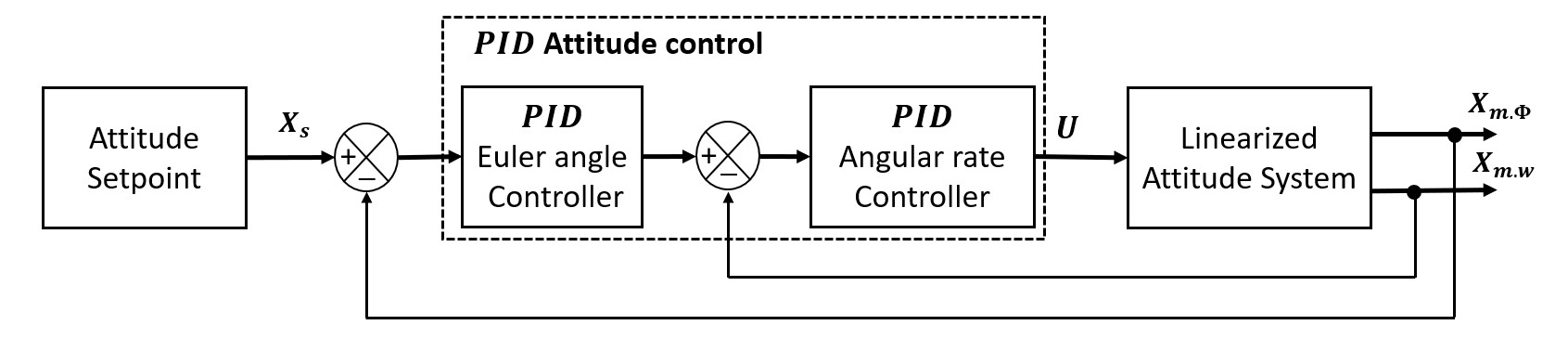}
\caption{Attitude control structure of UAVs using of PID control}
\label{Fig:PID_scheam}
\end{figure}

\indent PID (Proportional-Integral-derivative) control is a model-free control algorithm. 
A PID controller calculates an error value as the difference between the desired set point and measured point and then applies a correction based on a proportional, integral, and derivative terms as
\begin{align}\label{eqn:PID}
    \vo{u}(t) = {K_{P}}\hspace{0.1cm} \vo{e}(t) + {K_i} \int_{0}^{t} \vo{e}(t')dt' + {K_d} \frac{d e(t)}{dt}	
\end{align}
Most UAV systems currently use the PID controller for attitude control \cite{li2012survey}. 
Feedback measurement or estimated Euler angles and angular velocities \cite{kim2020nonlinear} are compared with the desired angle and angular velocity, respectively. 
The PID control generates an input value to eliminate the error. 
PID control framework for the attitude control is shown in Fig. \ref{Fig:PID_scheam}. 
For PID gain tuning, one can refer to \cite{bo2016quadrotor,wang2016dynamics,kim2020h2} for a more detailed analysis.

\subsection{LQR optimal control}
The linear quadratic regulator (LQR) is a method used in determining the state feedback controller $\vo{u} = \vo{K}_{LQR} \vo{x}$. This controller is designed to minimize the cost function, $J$, defined as 
\begin{align} \label{eqn:LQR_cost}
J = \int ^{\infty} _{0} (\vo{x}^T \vo{Q} \vo{x} + \vo{u}^T \vo{R} \vo{u}) dt  
\end{align}
where $\vo{Q}\geq \vo{0}$ and $\vo{R} > \vo{0}$ are symmetric weighting matrices.
These matrices are the main design parameters for defining the the control objective so that the state error and control energy is minimized. 
This cost function is solved with MATLAB function \texttt{lqr()}. 
The LQR problem can be converted to the LMI (Linear Matrix Inequality) form as given by the following theorem.
\begin{theorem}[\cite{duan2013lmis}]
The following two statements are equivalent:
\begin{enumerate}
    \item A solution $\vo{K}_{LQR}$ to the LQR controller exists.
    \item $\exists$ a matrix $\vo{Y}$, a symmetric matrix $\vo{W}$, and a symmetric matrix $\vo{Y}=\vo{P}^{-1}$ such that:
\end{enumerate}
\begin{align}
    \vo{AY}+ \vo{Y} \vo{A}^T +  \vo{W}^T \vo{B}_u ^T + \vo{B_{u}} \vo{W} &+ \vo{YQY} + \vo{W}^T \vo{RW} < 0  \label{LQR_LMI} 
\end{align}
The optimal LQR control gain, $\vo{K}_{LQR}$, is determined by solving the following optimization problem. 
\begin{align}
\min_{\vo{P},\vo{W},\vo{Y}} \quad  \textbf{trace}\  (\vo{P})  \quad \text{subject to (\ref{LQR_LMI}). \nonumber}
\end{align}
The gain $\vo{K}_{LQR}$ is recovered by $\vo{K}_{LQR} = \vo{W} \vo{Y}^{-1} $.
\end{theorem}
This optimal gain minimizes the cost function (\ref{eqn:LQR_cost}). 
To solve this optimized solution, we used CVX \cite{grant2009cvx} and MATLAB tool box \cite{gu2005robust}.

% \textbf{Theorem 1 (LQR Optimal Control)} \cite{duan2013lmis} : The following two statements are equivalent:
% \begin{enumerate}
%     \item A solution $\vo{K}_{LQR}$ to the LQR controller exists.
%     \item $\exists$ a matrix $\vo{Y}$, a symmetric matrix $\vo{W}$, and a symmetric matrix $\vo{Y}=\vo{P}^{-1}$ such that:
% \end{enumerate}
%         % \begin{align}\label{LQR_LMI} \nonumber
%         % (\vo{A}+\vo{BK})^T \vo{P} + \vo{P} (\vo{A}+\vo{BK}) \vo{Q} + \vo{K}^T \vo{R} \vo{K} < 0\\
%         % \textbf{trace}(\vo{P}) < \gamma^2
%         % \end{align}
% \begin{align}
%     % \vo{Y}=\vo{P}^{-1}, \nonumber \\
%     \vo{AY}+ \vo{Y} \vo{A}^T +  \vo{W}^T \vo{B}_u ^T + \vo{B_{u}} \vo{W} &+ \vo{YQY} + \vo{W}^T \vo{RW} < 0  \label{LQR_LMI} 
% \end{align}
% The optimal LQR control gain, $\vo{K}_{LQR}$, is determined by solving the following optimization problem. 
% \begin{align}
% \min_{\vo{P},\vo{W},\vo{Y}} \quad  \textbf{trace}\  (\vo{P})  \quad \text{subject to (\ref{LQR_LMI}). \nonumber}
% \end{align}
% The gain $\vo{K}_{LQR}$ is recovered by $\vo{K}_{LQR} = \vo{W} \vo{Y}^{-1} $. This optimal gain minimizes the cost function (\ref{eqn:LQR_cost}). To solve this optimized solution, please use cvx sofware \cite{grant2009cvx} and MATLAB tool box \cite{gu2005robust}. 

% \begin{align}
% \vo{e}(t) = \vo{x}(t) - \hat{\vo{x}} \rightarrow 0, \hspace{0.5cm} \text{ as } t \rightarrow \infty,
% \end{align}
%  and ensures that $\hat{\vo{x}}(t)$ is an asymptotic estimate of $\vo{x}(t)$.

\subsection{$\mathcal{H}_2$ Optimal Control}

With the linear system (\ref{eqn:lin_system}) and control law (\ref{eqn:control_law}), the $\mathcal{H}_2$ control closed-loop has the following form,
\begin{subequations} \label{control_loop}
\begin{align}
    \Dot{\vo{x}}(t) &= ( \vo{A} + \vo{B}_u \vo{K}) \vo{x}(t) + \vo{B}_z \vo{w}(t),\\
    \vo{z}(t) &= (\vo{C}_z + \vo{D}_u \vo{K} ) \vo{x}(t),
\end{align}
\end{subequations}
Therefore, the influence of the disturbance $\vo{w}$ on the output $\vo{z}$ is determined in frequency domain as $\vo{z} = \vo{G}_{z w}(s) \vo{w}(s)$
where $\vo{G}_{zw}(s)$ is the transfer function from the disturbance $\vo{w}$ to the output $\vo{z}$ given by
\begin{align} \label{errorTF}
\vo{G}_{z w}(s) = \vo{C}_z  (\vo{C}_z + \vo{D}_u \vo{K})[s \vo{I} -( \vo{A} + \vo{B}_u \vo{K})]^{-1} \vo{B}_w .
\end{align}

% where $\vo{K}$ is the control gain, and $\vo{z}(t)$ is the estimate of the output of interest. The error equations are then given by

% \begin{subequations} \label{estimator_Linear}
%     \begin{align}
%     \hat{\Dot{\vo{e}}}(t) &= (\vo{A}+\vo{L} \vo{C}_y) \hat{\vo{e}}(t) + (\vo{B}_w+\vo{L} \vo{D}_w)\vo{w}(t)\\
%     \Tilde{\vo{z}}(t) &= \vo{C}_z \hat{\vo{e}}(t)
%     \end{align}
% \end{subequations}

The problem of $\mathcal{H}_2$ optimal control design is then, given a system (\ref{errorTF}) and a positive scalar $\gamma$, find a matrix $\vo{K}=\vo{K}_{\mathcal{H}_2}$ such that
\begin{align}\label{min1}
\|\vo{G}_{zw}(s)\|_{2} < \gamma.
\end{align}
where $\|\vo{G}(.)\|_{2}$ is the corresponding 2-norm of the system.
The formulation to obtain $\vo{K}_{\mathcal{H}_2}$ is given by the following theorem.
\begin{theorem}[\cite{duan2013lmis, apkarian2001continuous,boyd1993control}]
The following two statements are equivalent:
\begin{enumerate}
    \item A solution $\vo{K}_{\mathcal{H}_{2}}$ to the $\mathcal{H}_2$ controller exists.
    \item $\exists$ a matrix $\vo{W}$, a symmetric matrix $\vo{Z}$, and a symmetric matrix $\vo{X}$ such that:
\end{enumerate}
\begin{align}\label{H2_LMI}
\vo{AX}+\vo{B}_u \vo{W} +(\vo{AX}+\vo{B}_u \vo{W})^T + \vo{B}_w \vo{B}_w ^T &< 0  \nonumber\\
\begin{bmatrix}\nonumber
\vo{-Z}  &  \vo{C}_z \vo{X} + \vo{D}_z \vo{W} \\
\vo{*} & \vo{-X}
\end{bmatrix} &< 0\\
\textbf{trace}(\vo{Z}) &< \gamma^2
\end{align}
The minimal attenuation level $\gamma$ is determined by solving the following optimization problem
\begin{align}
\min_{\vo{W},\vo{X}, \vo{Z}} \gamma \quad \text{ subject to (\ref{H2_LMI}). }  \nonumber
\end{align}
The $\mathcal{H}_2$ optimal control gain is recovered by $\vo{K}_{\mathcal{H}_2}= \vo{W} \vo{X}^{-1}$.
\end{theorem}
 
\begin{figure}[ht] 
\centering
\includegraphics[width=0.48\textwidth]{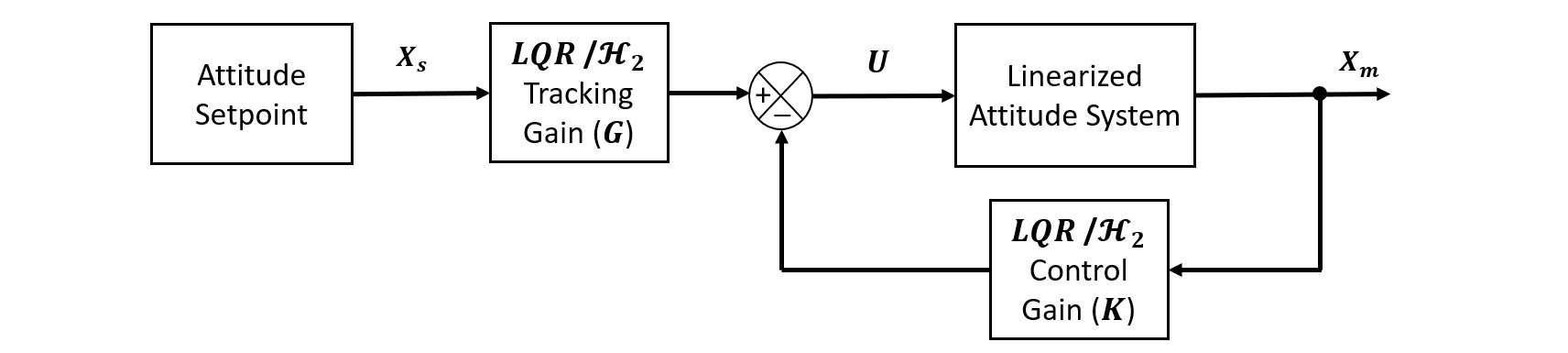}
\caption{Attitude control structure of UAVs using  $LQR$ and $\mathcal{H}_2$ controller}
\label{Fig:H2_scheam}
\end{figure}

This optimal gain ensures that the closed-loop system is asymptotically stable and attenuates the disturbance. % since this include the disturbance in the cost function. 
To solve this optimization problem, we use CVX \cite{grant2009cvx} and Matlab tool box \cite{gu2005robust}. $LQR$ and $\mathcal{H}_2$ control framework for the attitude control is shown in Fig. \ref{Fig:H2_scheam}.

\section{Results}\label{section:result}

\subsection{Simulation set up}

\begin{figure}[ht]
\centering
    \includegraphics[trim=0.8cm 0.3cm 1cm 0.5cm,clip,width=0.22\textwidth,height=2cm]{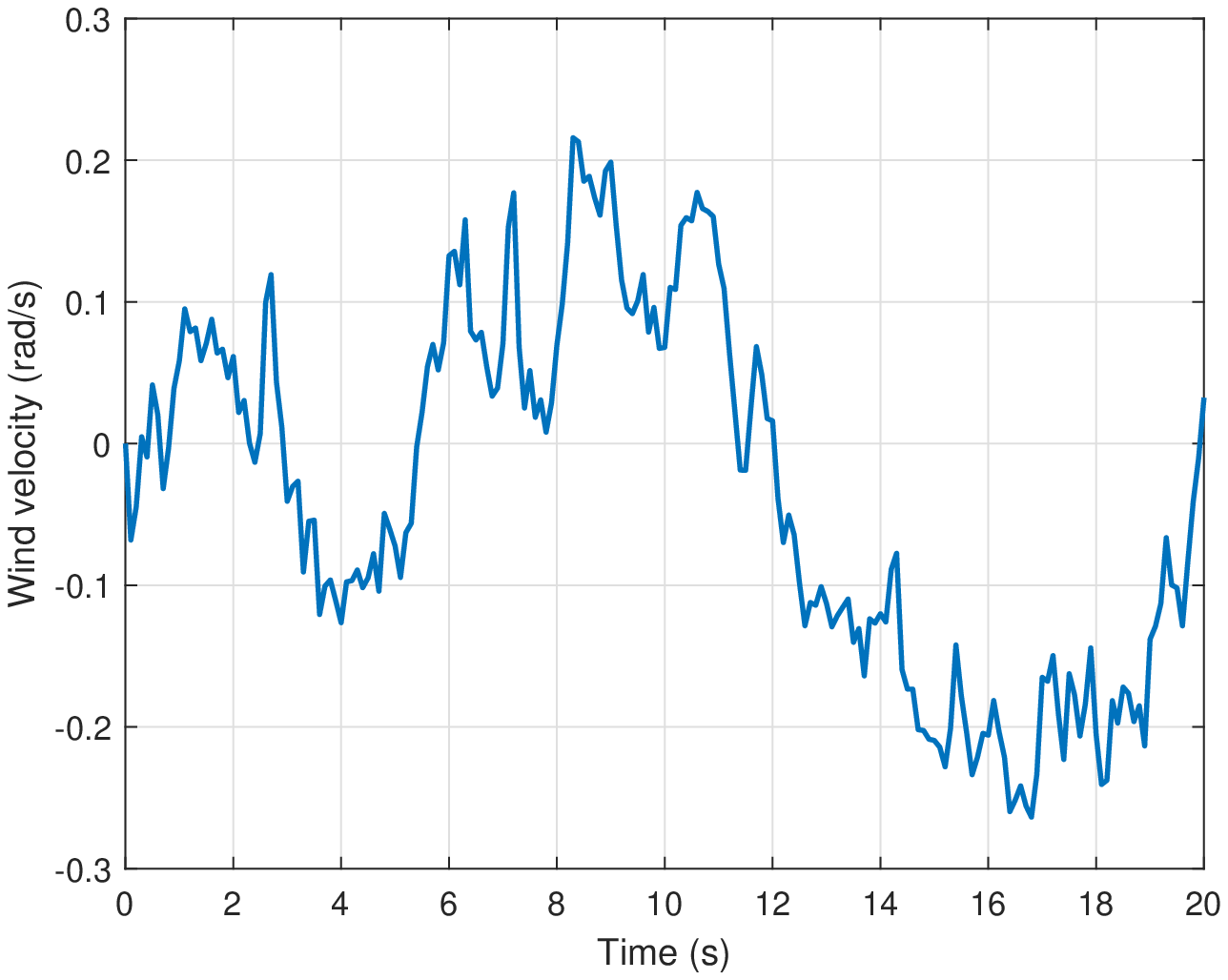} \quad
    \includegraphics[trim=0.5cm 0.3cm 1cm 0.5cm,clip,width=0.22\textwidth,height=2cm]{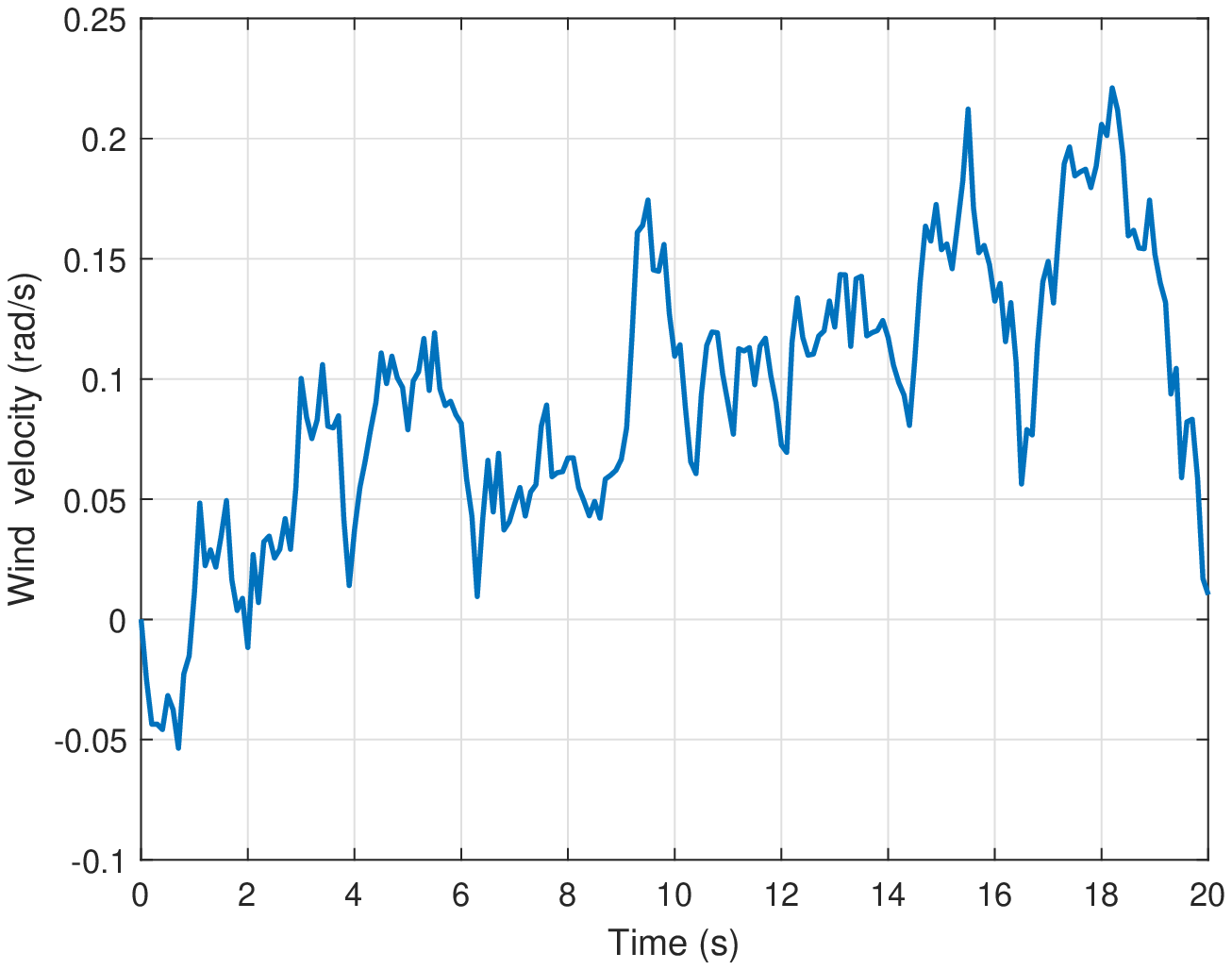} \quad
    \caption{Angular velocity component of wind disturbance about the X (Left) and Y (Right) axis generated by the Dryden wind turbulence model in the Simulink software. }
    % \comment {* Z component is not shown in this plot.}}
\label{Fig:wind}
\end{figure}

\begin{figure}[ht] 
\centering
\includegraphics[width=0.48\textwidth]{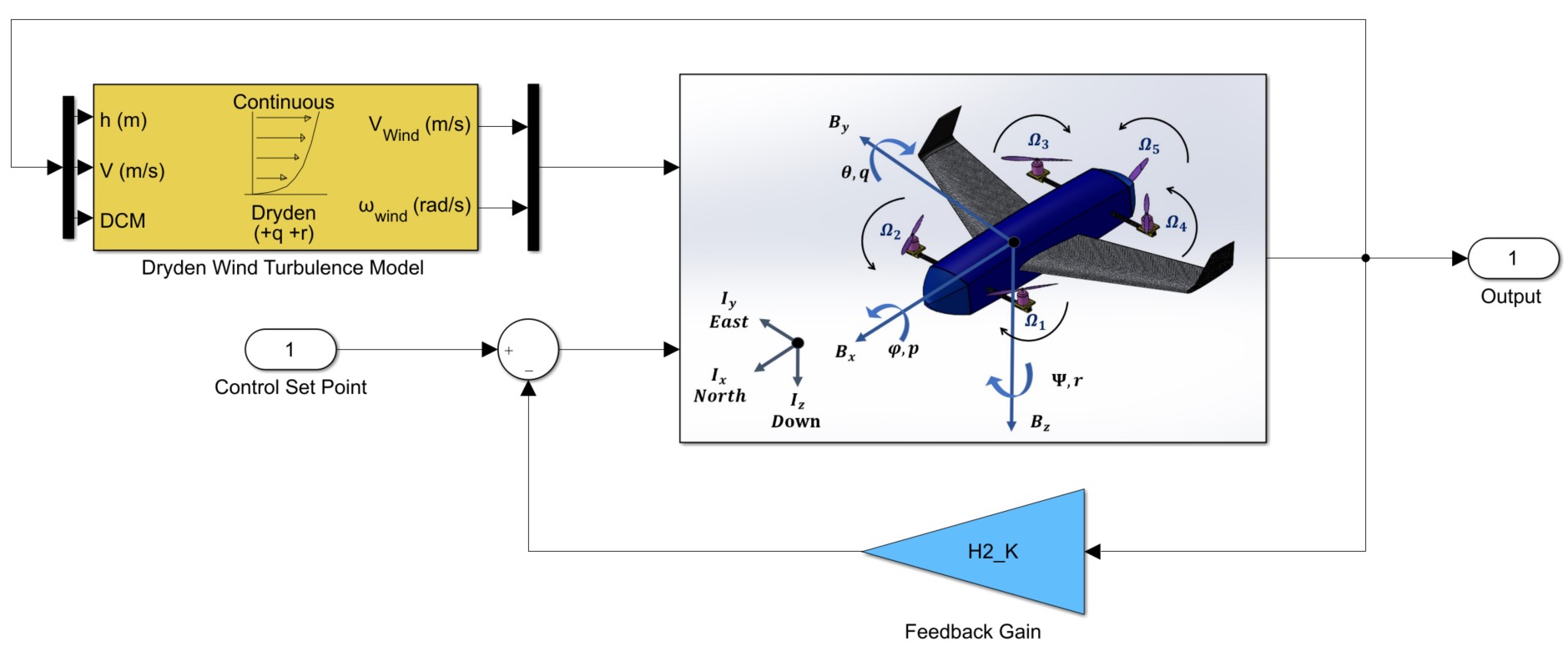}
\caption{6-DOF non-linear simulation of the hybrid UAV with disturbance}
\label{Fig:Non_Sim}
\end{figure}

The proposed  $\mathcal{H}_2$ optimal control is applied to attitude control of the linearized dynamics of our UAV as modeled by Eq. \ref{eqn}. 
We compare its performance with the PID controller and LQR. 
The comparison is done with respect to the control input, system response, and the amount of wind disturbance rejection, in a Simulink based simulation environment, as shown in Fig. \ref{Fig:Non_Sim}. 
In this simulation, the Dryden wind turbulence model was used to generate the wind disturbance. 
The generated wind disturbance is 10 m/s from north. Angular velocity components of the wind along X and Y axes are shown in Fig. \ref{Fig:wind}. 
The final simulation environment which includes the UAV system, controller, and disturbance model is shown in Fig. \ref{Fig:Non_Sim}.

% The noise in the motor is generated as torque values with normal distribution \comment{what is the mean and covariance and how did you choose them, any reference} 

% and imported in simulink software. 

% \begin{figure}[h]
% \centering
% % \includegraphics[scale=0.27]{image/Vertical_control1.jpg}
% \includegraphics[width=0.4\textwidth]{image/Vertical_control1.jpg}
% \caption{Simulation structure for vertical altitude control}
% \label{Fig:simulation}
% \end{figure}

% The Figure \ref{Fig:Non_Sim} is our final non-liner simulation which includes the UAV system, controller, and disturbance model. We will test our control and estimation algorithm with this simulation.  

We simulated two cases: \textit{Case I -- Level flight (Longitudinal motion)} which considers parameters in Eq. (\ref{eqn:sys_level}) for level flight trim states in Eq. (\ref{eqn:trim_level}).  
Input of the system is deflection angle of elevon surface and measurement is angular velocity $q$. 
Initial deviation of angular velocity about Y axis in body frame $p$, is $0.5$ rad/sec.
\textit{Case II-- Hover flight}  which consider parameters in Eq. (\ref{eqn:sys_hover}) for hover at trim states in Eq. (\ref{eqn:trim_hover}). 
Input to the system is the PWM signals of four motors and, measurement are all state, Euler angle and angular velocity. 
Initial deviation of pitch angle $\theta$, is $10^\circ$.

LQR (\ref{LQR_LMI}), PID (\ref{eqn:PID}), and $\mathcal{H}_2$ (\ref{H2_LMI}) controllers are designed with these two linearized systems and then tested in the non-linear model in Fig \ref{Fig:Non_Sim}.

\subsection{Simulation results}
We examine the performance of the $\mathcal{H}_2$ control by comparing it with that of the PID controller and LQR in terms of root mean squared (RMS) error and time response.

\begin{figure}[ht] 
\centering
\includegraphics[trim=0cm 1cm 0cm 0.5cm,width=0.48\textwidth]{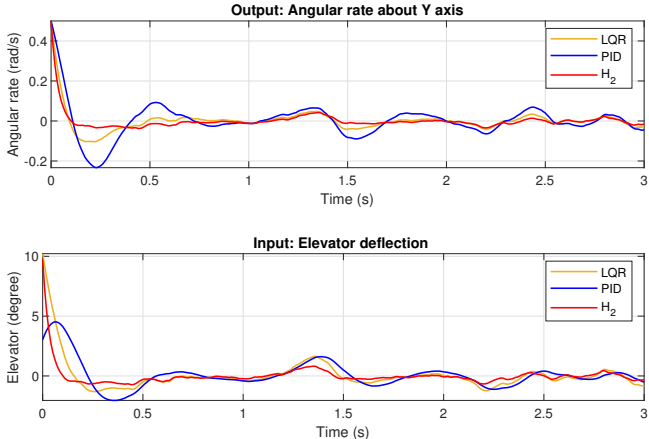}
\caption{Error comparison of LQR, PID, and $\mathcal{H}_2$ control with wind disturbance in level flight}
\label{Fig:level_out}
\end{figure}

\textbf{Case I: Level flight} -- The simulation results for the proposed $\mathcal{H}_2$ control, the PID, and the LQR are shown in  Fig. \ref{Fig:level_out} and TABLE \ref{Table:RMS_levelFlight}.
The proposed $\mathcal{H}_2$ control has the least RMS error than the other controllers, as shown in TABLE \ref{Table:RMS_levelFlight}. 
The time response and overshoot of $\mathcal{H}_2$ control is noted to be shorter than one of the PID controller and the LQR. 
% \vspace{-0.5cm}

\begin{table}[h]
\caption{RMS error for level flight: Case I.} \label{Table:RMS_levelFlight}
\vspace{-0.5cm}
\begin{center}
\renewcommand{\arraystretch}{1.5}
\begin{tabular}{|c||c|c|c|}
\hline
Algorithm & $LQR$ & $PID$ & $\mathcal{H}_2$\\
\hline \hline
q ($rad/sec$) & 0.0573 &  0.0859 & 0.0457\\
\hline
\end{tabular}
\end{center}
\end{table}

\begin{figure}[ht] 
\centering
\includegraphics[trim=0cm 0.5cm 0cm 0.3cm,clip,width=0.48\textwidth]{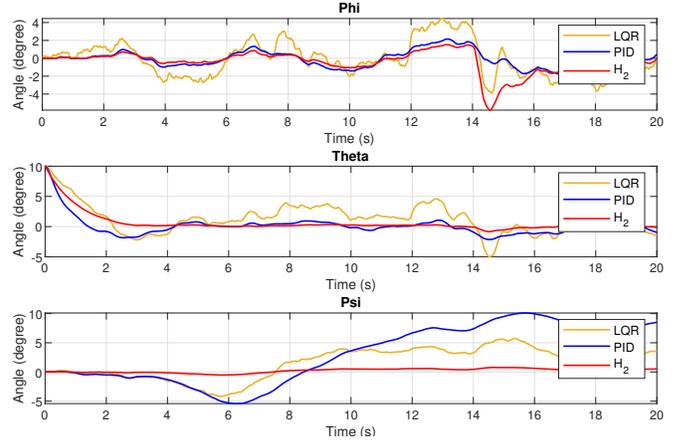}
\caption{Error comparison of LQR, PID, and $\mathcal{H}_2$ control with wind disturbance in Hover flight}
\label{Fig:hover_out}
\end{figure}

\textbf{Case II: Hover flight} --  The simulation results for the proposed $\mathcal{H}_2$ control, PID, and the LQR are shown in  Fig. \ref{Fig:hover_out} and TABLE \ref{Table:RMS_hoverFlight}.
The proposed $\mathcal{H}_2$ control has the least RMS error compared to the other controllers, as shown in TABLE \ref{Table:RMS_hoverFlight}, especially in yaw angle ($\psi$). 
The time response of proposed $\mathcal{H}_2$ control is comparable with one from the PID controller and LQR. 
Here, note that $\mathcal{H}_2$ is implicitly a better algorithm to deal with disturbance since it include disturbance as a design factor.

\vspace{-0.5cm}
\begin{figure}[ht] 
\centering
\includegraphics[trim=0cm 0.3cm 0cm .5cm,clip,width=0.48\textwidth]{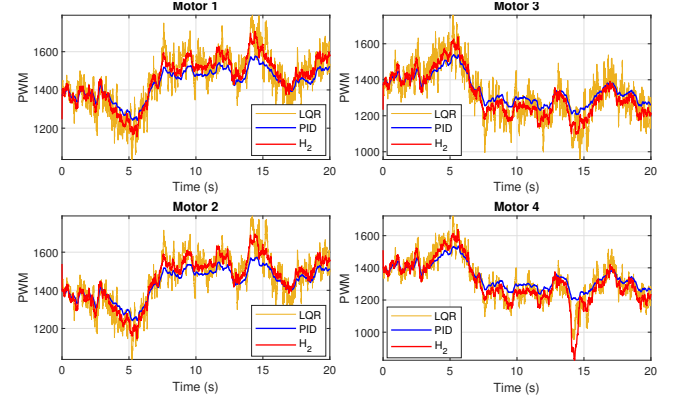}
\caption{Input comparison of LQR, PID, and $\mathcal{H}_2$ control with wind disturbance in hover flight}
\label{Fig:hover_in}
\end{figure}

\begin{table}[h]
\caption{RMS error for the hover flight: Case II.} \label{Table:RMS_hoverFlight}
\vspace{-0.5cm}
\begin{center}
\renewcommand{\arraystretch}{1.5}
\begin{tabular}{|c||c|c|c|}
\hline
Algorithm & Roll angle (${}^\circ$) & Pitch angle (${}^\circ$) & Yaw angle (${}^\circ$)\\
\hline \hline
$LQR$  & 0.8964 & 1.9441 & 3.0217\\
\hline
 $PID$& 0.0349 & 1.3169 & 5.7745\\
\hline
 $\mathcal{H}_2$ & 0.1878 & 1.5935 & 0.4370\\
\hline
\end{tabular}
\end{center}
\end{table}

\section{Conclusion}
\label{section:conclusion}
This paper presents an approach to design a vertical take-off and landing hybrid UAV. 
We elaborately describe its modeling and controller design that will make it robust to wind disturbances. We discuss methods that rapidly implements the modeling of our proposed hybrid UAV satisfying the requirements with sufficient accuracy.
We also propose a robust controller based on $\mathcal{H}_2$ optimal theory for our hybrid UAV. 
This controller achieves better performance while rejecting wind gusts compared to that of the PID and the LQR controller. For the future work, discrete time system of UAV  will be developed and tested in physical UAV model.

% or


% \appendices

% % you can choose not to have a title for an appendix
% % if you want by leaving the argument blank
% \section{}

% % use section* for acknowledgement
% \section*{Acknowledgment}


% % Can use something like this to put references on a page
% % by themselves when using endfloat and the captionsoff option.
% \ifCLASSOPTIONcaptionsoff
%   \newpage
% \fi



% \bibliographystyle{IEEEtran}
\bibliography{Hybrid}
\bibliographystyle{unsrt}
\end{document}